\documentclass[fleqn,usenatbib]{mnras}

\usepackage[T1]{fontenc}

\DeclareRobustCommand{\VAN}[3]{#2}
\let\VANthebibliography\thebibliography
\def\thebibliography{\DeclareRobustCommand{\VAN}[3]{##3}\VANthebibliography}

\usepackage{graphicx}	
\usepackage{amsmath}	
\usepackage{amssymb}	
\usepackage{tabularx}

\usepackage{caption}
\usepackage{subcaption}

\graphicspath{{images/}}

\title[The broadening of the main sequence in M38]{The broadening of the main sequence in the open cluster M38}

\author[M. Griggio et al.]{
M. Griggio,$^{1,2}$\thanks{E-mail: massimo.griggio@inaf.it}
M. Salaris,$^{3,4}$
L. R. Bedin,$^{2}$
and S. Cassisi$^{4,5}$
\\
$^{1}$Dipartimento di Fisica, Universit\`a di Ferrara, Via Giuseppe Saragat 1, Ferrara I-44122, Italy\\
$^{2}$INAF - Osservatorio Astronomico di Padova, Vicolo dell'Osservatorio 5, Padova I-35122, Italy\\
$^{3}$Astrophysics Research Institute, Liverpool John Moores University, 146 Brownlow Hill, Liverpool L3 5RF, UK\\
$^{4}$INAF - Osservatorio Astronomico di Abruzzo, Via M. Maggini, I-64100 Teramo, Italy\\
$^{5}$INFN - Sezione di Pisa, Largo Pontecorvo 3, 56127 Pisa, Italy
}

\date{Accepted 2023 June 07. Received 2023 June 07; in original form 2023 April 26}

\pubyear{2023}

\usepackage{newtxtext,newtxmath}

\begin{document}
\label{firstpage}
\pagerange{\pageref{firstpage}--\pageref{lastpage}}
\maketitle

\begin{abstract}
Our recent multi-band photometric study of the colour width of the lower main sequence of the open cluster M37 has revealed the presence of a sizeable initial chemical composition spread in the cluster.
If initial chemical composition spreads are common amongst open clusters, this would have major implications for cluster formation models and the foundation  of the chemical tagging technique.  
Here we present a study of the unevolved main sequence of the open cluster M38, employing 
{\it Gaia} DR3 photometry and astrometry, together with newly acquired {\it Sloan} photometry. We have analysed the distribution of the 
cluster's lower main sequence stars with a differential colour-colour diagram made of combinations of {\it Gaia} and {\it Sloan} magnitudes, like 
in the study of M37.
We employed synthetic stellar populations to reproduce the observed trend of M38 stars in this diagram, and found that   
the observed colour spreads can be explained simply by the combined effect of differential reddening across the face of the cluster and 
the presence of unresolved binaries.  
There is no need to include in the synthetic sample a spread of initial chemical composition as instead necessary to explain the main sequence of M37.
Further photometric investigations like ours, as well as accurate differential spectroscopic analyses on large samples  
of open clusters, are necessary to understand whether chemical abundance spreads are common among the open cluster population.

\end{abstract}




\section{Introduction}

Open clusters have been traditionally considered to host populations of stars born all with the same initial chemical composition in a burst of star formation of negligible duration (simple stellar populations).

The recent discovery of extended turn offs (TOs) in the {\it Gaia} colour-magnitude diagrams (CMDs) of a sample of about 15 open clusters with ages in the range $\sim$\,0.2-1\,Gyr and initial metal mass fractions $Z$ between $\sim$\,0.01 and $\sim$\,0.03 
\citep{marino18, bastian, cordoni18} has somehow challenged this paradigm, given that extended TOs can be naturally explained by a range of ages amongst the cluster's stars \citep[e.g.][]{mb, corr}.
Further detailed studies of the extended TO phenomenon, which is seen also in CMDs of Magellanic Clouds' clusters younger than 2\,Gyr \citep[see, e.g.,][and references therein]{mackey,pb16,goudrot}, strongly point to the effect of rotation 
\citep[e.g.][]{bastian, kamann18, kamann20, kamann23}
as the main culprit \citep[see also][]{dantona}.
In this case, stellar populations in individual open clusters might still be simple stellar populations, born with uniform age and initial chemical composition.

Very recently, our photometric multi-band study of the main sequence (MS) of the open cluster M37 \citep{griggiospread} has disclosed the presence of a sizeable initial chemical composition spread in the cluster (either a full metallicity range $\Delta\rm{[Fe/H]}$\,$\sim$\,0.15\,dex or a helium mass fraction total range $\Delta Y$\,$\sim$\,0.10). This result is independent of whether rotation or age spread is responsible for its observed extended TO, because it is based on an analysis of the lower MS, populated by stars with convective envelopes that are anyway slow rotators.

This result has important implications for our understanding of open cluster formation \citep[see, e.g.][]{clarke} and 
the technique of \lq{chemical tagging\rq} of Galactic field stars \citep[e.g.,][]{tagging, hogg}, especially if high resolution spectroscopic investigations of M37 will disclose that the chemical spread is due to an inhomogeneous initial metal content. 
Indeed, the basic idea of chemical tagging is that stars are born in unbound associations or star clusters (like open clusters) that disperse rapidly, and over time they populate very different parts of the Milky Way phase space; stars of common birth origin should however be identifiable through their measured photospheric abundances, in the assumption that their birth cluster has a chemically homogeneous composition.
It is therefore important to assess whether initial abundance spreads among the Galactic open clusters are a common phenomenon.

In this paper we have investigated the poorly-studied open cluster M38 that, like M37, displays in the \textit{Gaia} Data Release 3
\citep[DR3,][]{2022arXiv220800211G} CMD a MS broader than what is expected from photometric errors only. 
We have applied the same multi-band technique developed for M37 that combines both {\it Gaia} and {\it Sloan} photometry, to assess whether 
the broadening of the MS can be explained by differential reddening and binaries only, or whether a chemical abundance spread is also required.

The plan of the paper is as follows. Section \ref{Gaia} presents our membership analysis and the resulting \textit{Gaia} DR3 CMD, and is 
followed by Section\,\ref{Sloan} which describes the complementary {\it Sloan} photometry used in this work. 
Section\,\ref{broad} describes the theoretical analysis of the MS width 
and Section\,\ref{conclusions} closes the paper with our conclusions.


\section{The \textit{Gaia} colour-magnitude diagram}
\label{Gaia}

\begin{figure}
    \centering
    \includegraphics[width=\columnwidth]{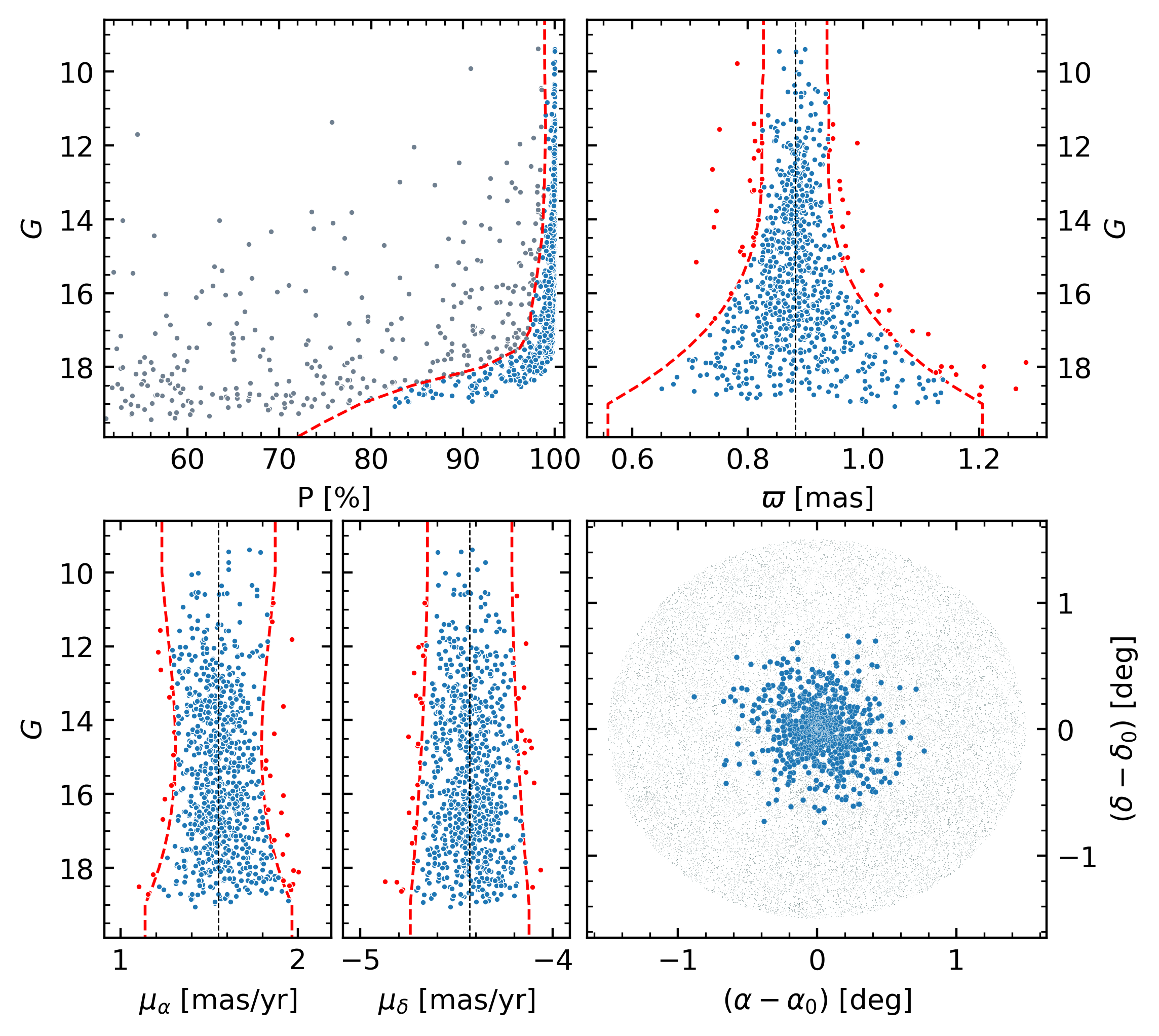}
    \caption{Cluster member selection. {\sl Top left panel:} Membership probability for all the sources. In our analysis we reject stars with $P<50\,\%$.
    {\sl Top right panel:} $G$ magnitude vs parallax. In this case we reject stars that fall outside the region bounded by the two red lines (red dots).
    The black dashed line denotes the median parallax. {\sl Bottom left panel:} Proper motions vs $G$ magnitude of all sources that passed the two 
    previous selections. In this case we kept the sources lying between the red dashed lines. {\sl Bottom right panel:} spatial distribution of the sources.
    The blue dots denote our selected members of M38.}
    \label{fig:mp}
\end{figure}

\begin{figure}
    \centering
    \includegraphics[width=\columnwidth]{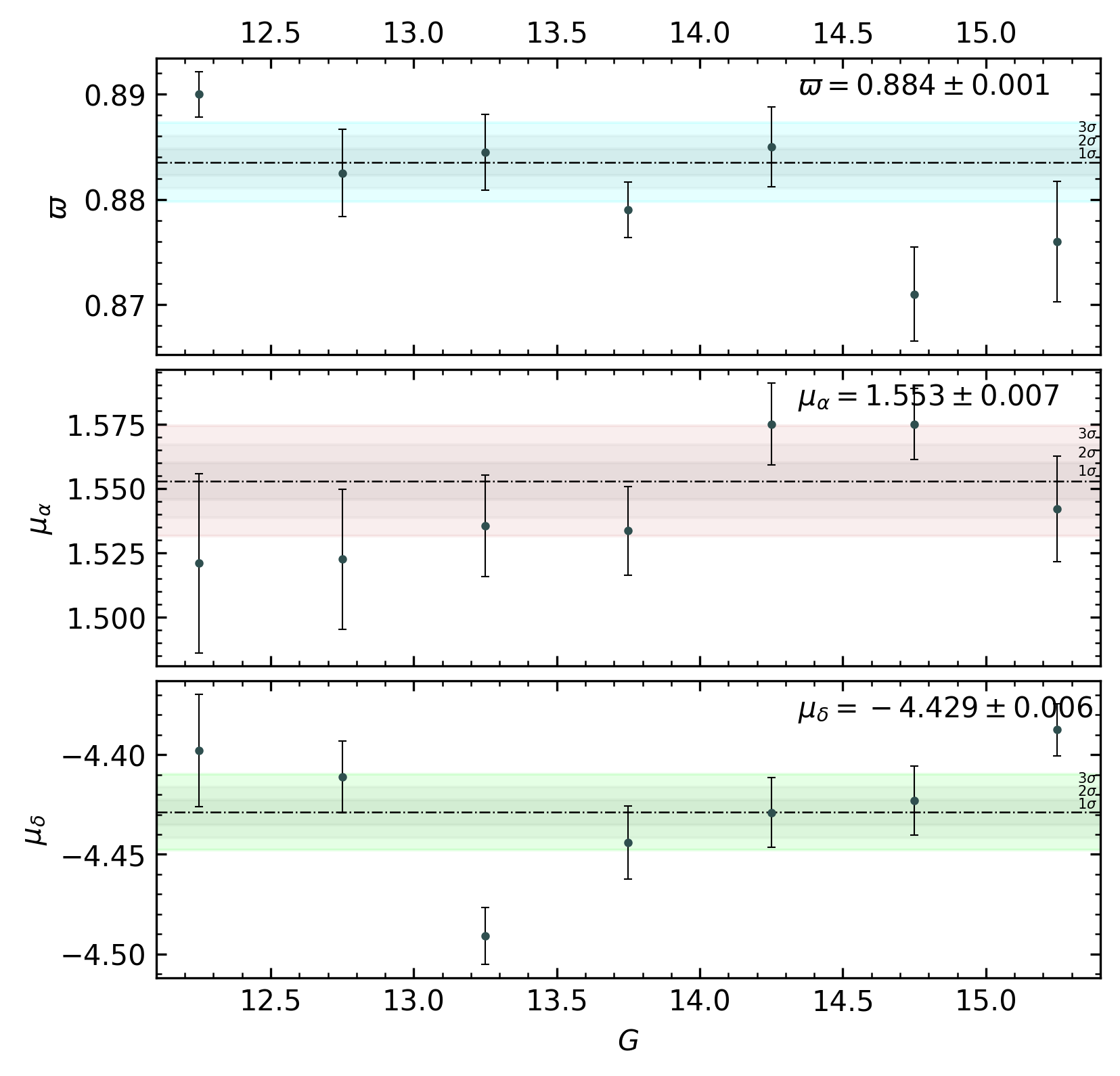}
    \caption{Mean values of the parallax and the components of the proper motion in each magnitude bin;
    the dash-dotted line denotes the overall weighted mean values, also reported at the top right corner of each panel
    (in mas and mas\,yr$^{-1}$ respectively.) and in Table\,\ref{tab:params}.}
    \label{fig:par}
\end{figure}

The analysis of the CMD diagram of a star cluster requires a sample of member stars free from
field sources contamination. To obtain such a sample we have derived the membership probabilities for
all the sources in the {\it Gaia} DR3 catalogue within a circle with a $\sim$\,1.5\,deg radius,
centred on the cluster \citep[$\alpha_0=82.167,\,\delta_0=35.824$,][]{tarricq}.

The membership probabilities were computed following the approach described by 
\cite{gb2022}, which relies on {\it Gaia} DR3 astrometry. Cluster members were
selected by performing a series of cuts on the astrometric parameters, as displayed
in Fig.\,\ref{fig:mp}. In the top left panel we show the membership probability
$P$: we applied a cut by-eye, following the profile of the bulk of sources with cluster
membership at each magnitude (dashed-red curve). This selection becomes less strict
at fainter magnitudes, as the measurement errors increase and memberships
become less certain.
We then applied a cut on the parallax
(top right) and proper motions (bottom left) distributions. The red lines were
defined by the 68.27$^{\rm th}$ percentile of the residuals around their median value 
in each 1-magnitude bin, multiplied by a factor of two \citep[as in][]{griggio22}.
The bottom right panel show the spatial distribution of the selected members.\\

The derived list of probable cluster members allowed us to estimate the
cluster astrometric parameters; we followed the same procedure as in
\cite{gb2022}, by applying some quality cuts to the {\it Gaia} data, i.e.:

\begin{itemize}
    \item[-] $\texttt{astrometric\_excess\_noise}<0.25$;
    \item[-] $\texttt{phot\_bp\_rp\_excess\_factor}<1.4$;
    \item[-] $\texttt{phot\_proc\_mode}=0$;
    \item[-] $\texttt{astrometric\_gof\_al}<4$;
    \item[-] $\sigma_\varpi/\varpi<0.1$, $\sigma_{\mu_\alpha}/\mu_\alpha<0.1$ and $\sigma_{\mu_\delta}/\mu_\delta<0.1$.
\end{itemize}

With this selected sample of members we have estimated the cluster's mean proper
motion and parallax. The mean values in each magnitude interval are shown
in Fig.\,\ref{fig:par}, with the weighted average reported on the top right corner of each panel.
The cluster parameters are also reported in Table\,\ref{tab:params}

\begin{table}
    \centering
    \caption{M38's astrometric parameters estimated in this work.}
    \label{tab:params}
    \begin{tabularx}{\columnwidth}{XXXX}
        \hline \hline
        $\mu_\alpha$\,[mas\,yr$^{-1}$] & $\mu_\delta$\,[mas\,yr$^{-1}$] & $\varpi$\,[mas] & Distance\,[pc] \\ \hline
        $1.553\pm0.007$ & $-4.429\pm0.006$ & $0.884\pm0.001$ & $1130\pm50$ \\
        \hline
    \end{tabularx}
\end{table}

The average parallax gives a distance $d$ of $1132\pm2$\,pc, that, 
accounting for the
parallax zero-point correction by \cite{2021A&A...649A...4L}, becomes $1183\pm2$\,pc.
In the following, we consider this correction to represent a maximum error in the distance, hence  
$d=1130\pm50$\,pc. Our estimate is also in agreement, within the errors, with the distances given by \cite{2021AJ....161..147B} which provide a median value for M38 stars equal to $1186\pm2$\,pc.

\begin{figure}
    \centering
    \includegraphics[width=\columnwidth]{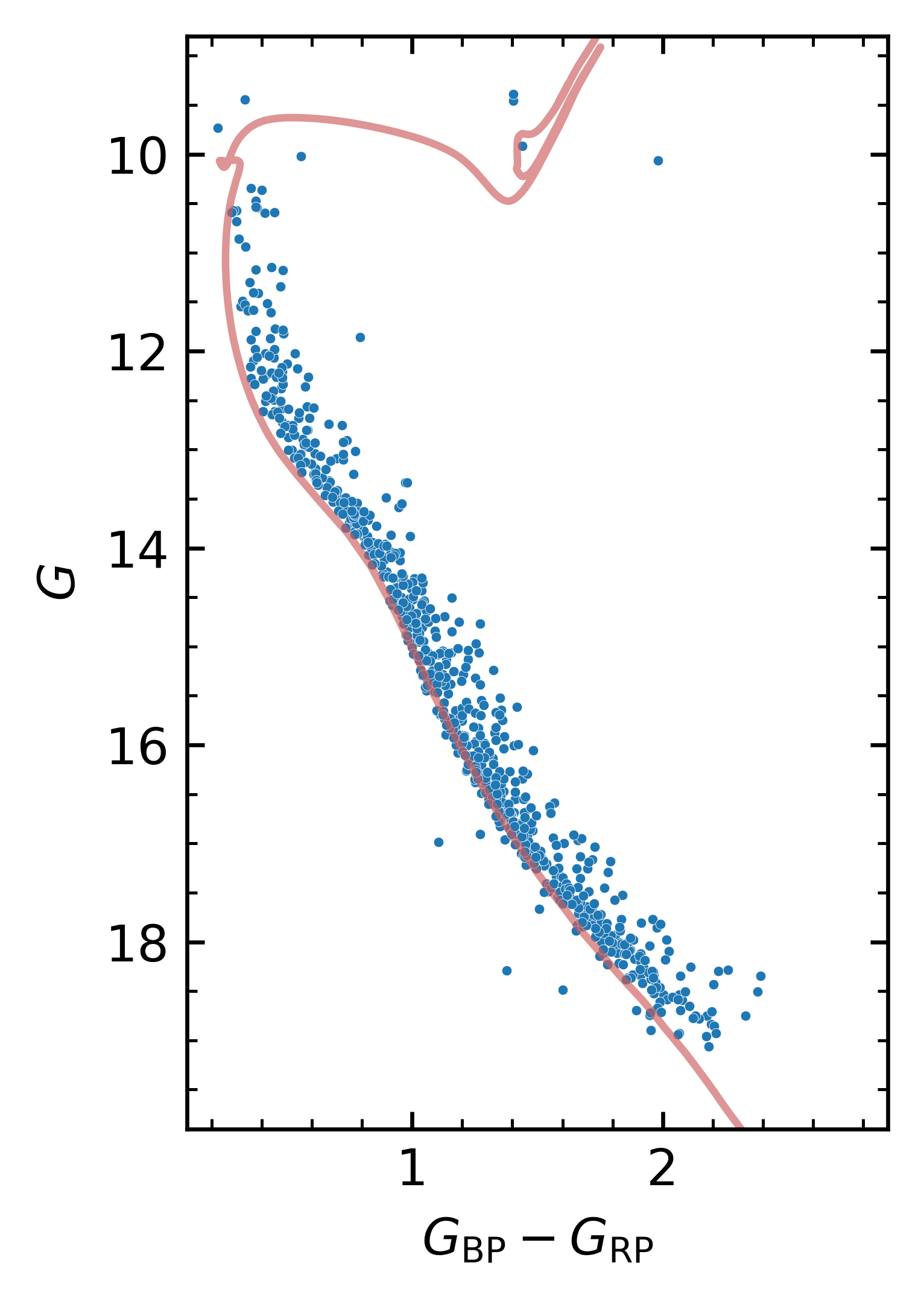}
    \caption{{\it Gaia} CMD for the selected members of M38.
    The red line is a 300\,Myr BaSTI-IAC isochrone, with $\rm[Fe/H]=0.06$, shifted by $E(B-V)=0.26$ and $d=1130$\,pc (see text for details).}
    \label{fig:cmd_gaia}
\end{figure}

The CMD of the selected cluster members is shown in Fig.\,\ref{fig:cmd_gaia}. 
The MS is very well-defined and does not exhibit a clear extended TO. 
However, a detailed analysis of the TO region is hampered by the fact that there are only 40 MS stars with $G>12$.

The metallicity of this cluster is not well determined, given that spectroscopic analyses of small samples of cluster stars have provided a range of [Fe/H] determinations  
between $\sim$\,$-0.07$ and $\sim$\,$-0.38$, and  
$E(B-V)$ estimates range between $\sim$\,0.25 and $\sim$\,0.35\,mag \citep[see, e.g.,][]{subramaniam, dias, majaess, frincha, carrera, donor, zhong, li}.

In the same Fig.\,\ref{fig:cmd_gaia} we show for reference 
a 300\,Myr\footnote{This value is consistent with the age of 302\,Myr assigned to this cluster by \citet{tarricq}.} BaSTI-IAC \citep{hidalgo18} solar scaled isochrone with $\rm [Fe/H]=0.06$, matched to the 
blue edge of the lower MS (see below and Sect.\,\ref{broad} for the definition of lower MS and its blue edge). We adopted
the distance $d=1130$\,pc, and for the assumed metallicity we determined $E(B-V)=0.26$ from the match to the lower MS colour, which can be considered to be the minimum value of the 
reddening, given the presence of differential reddening across the face of the cluster, as discussed later in Sect.\,\ref{broad}.
We employed extinction coefficients 
in the {\it Gaia} bands obtained from the relations given in the {\it Gaia}
website\footnote{\url{https://www.cosmos.esa.int/web/gaia/edr3-extinction-law}}.

We tried also isochrones with lower [Fe/H] more in line with the uncertain spectroscopic estimates ($\rm [Fe/H]=-0.08$ and $-0.20$\,dex). After adjusting (actually increasing) $E(B-V)$ to match the blue edge of the lower MS, and the isochrone age to approximately reproduce the observed brightness of the TO region, the fit of the upper MS was poorer when considering these subsolar metallicities. 

We stress at this stage that --as for the case of M37-- 
the results of the analysis in Sect.\,\ref{broad} are insensitive to the exact values of the adopted isochrone metallicity, the cluster distance (within the adopted error bar), and the minimum value of $E(B-V)$, because of the differential nature of the technique applied.

\subsection{The width of the MS}

As discussed for M37 \citep{griggiospread}, if open clusters host single-metallicity populations, 
the observed colour width of the unevolved MS is expected to be set by 
the photometric error, the presence of unresolved binaries with a range of values of the mass ratio $q$, and the differential reddening across 
the face of the cluster, if any.
To verify this hypothesis in the case of M38, we have followed the same procedure 
detailed in \citet{griggiospread}.

\begin{figure}
    \centering
    \includegraphics[width=.95\columnwidth]{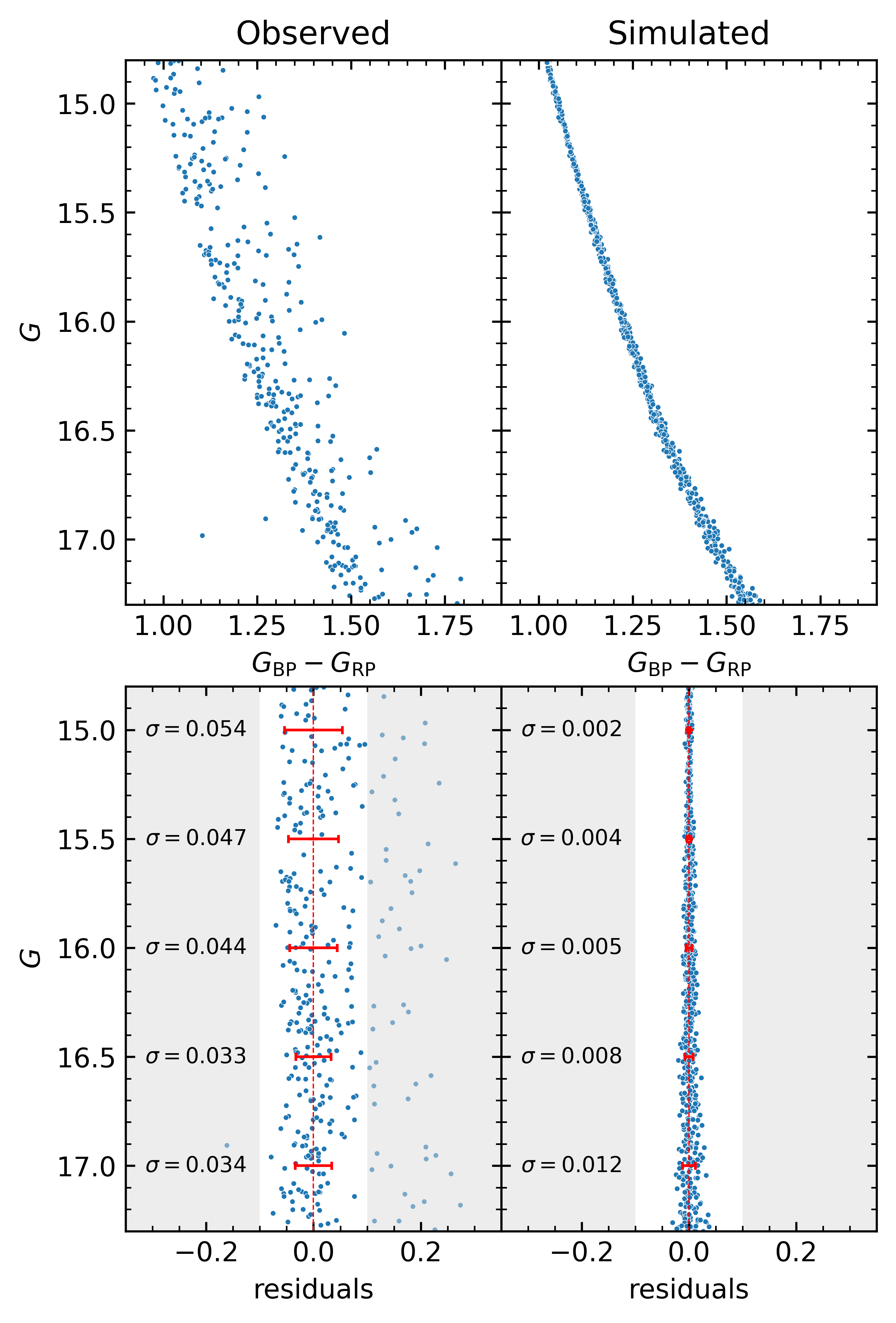}
    \caption{\textit{Top panels}: Observed (left) and simulated (right) CMDs of the lower MS of M38. \textit{Bottom panels}: Colour residuals as a function of the $G$ magnitude around the 
    observed (left) and the synthetic (right) CMD fiducials. The area shown in white contains the stars employed to compute
    the $1\sigma$ values of the dispersion of the residuals also shown 
    in the two panels (see text for details).}
    \label{fig:cmd_err}
\end{figure}

In brief, we first calculated an observed fiducial line 
of the unevolved MS in 
the $G$-magnitude range between 15.2 and 16.6 (denoted as lower-MS from now on). According to the isochrone 
in Fig.\,\ref{fig:cmd_gaia}, in this magnitude range the single star population covers a mass range between $\sim$\,0.9 and 1.15\,$M_{\odot}$, approximately the same range as in our analysis of the lower MS of M37 \citep[see][]{griggiospread}.
We have calculated the fiducial line assuming that the observed MS is populated just by single stars all with the same initial metallicity, as 
described in \citet{griggiospread}.
Synthetic stars have been then distributed with uniform probability along this fiducial; each synthetic magnitude 
has been then perturbed by a 
photometric error obtained by randomly sampling a Gaussian probability distribution with zero mean and a standard deviation set to the median error at the corresponding 
$G$-magnitude, taking advantage of the individual errors from the {\it Gaia} DR3 catalogue. 
Figure\,\ref{fig:cmd_err} (top panels) compares the observed CMD (left) with the simulated counterpart (right) in the selected magnitude range, and the colour residuals around the fiducial line 
as a function of $G$ (bottom panels).
We also show the values of the colour dispersion around the fiducial values at varying magnitudes in both CMDs. They have been computed as the $68.27^{\rm th}$-percentile of the distribution of the residuals around zero.

Notice that we have discarded objects with the position in the CMD consistent with being  
unresolved binaries with mass ratio $q > 0.7$ (according to the adopted isochrone), when we calculated the dispersion of the residuals from the observations. But 
even neglecting these objects, it is clear from Fig.\,\ref{fig:cmd_err} that the simulated stars display a much narrower distribution around the fiducial line than the observations.

To assess the origin of the colour spread of the observed CMD, we employed an auxiliary photometry in the {\it Sloan} $ugi$ filters --described in the following section-- and applied in Sect.\,\ref{broad} the same technique developed in \citet{griggiospread}.

\section{\textit{Sloan} observations and data reduction}
\label{Sloan}

\begin{figure}
    \centering
    \includegraphics[width=.95\columnwidth]{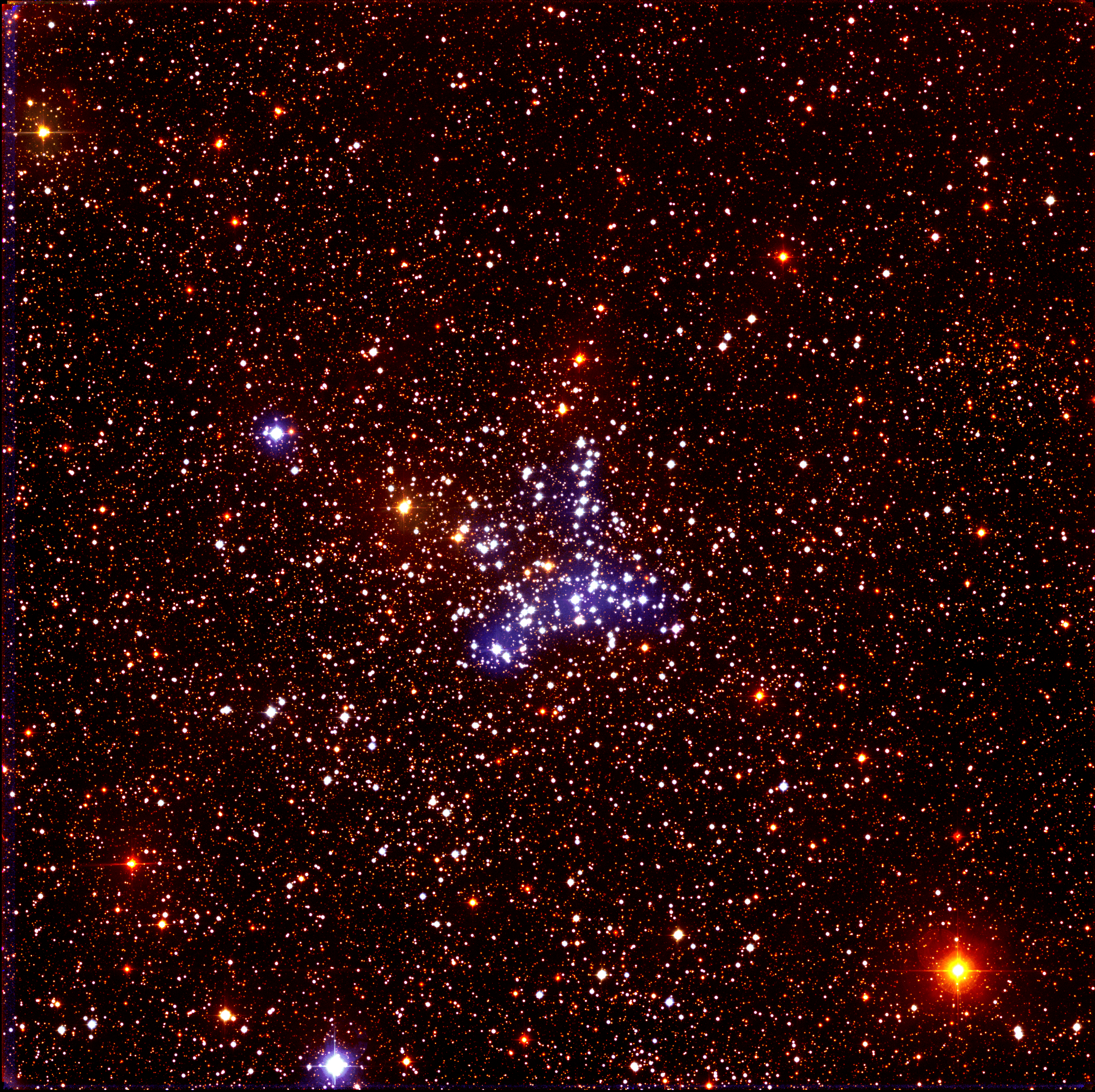}
    \caption{Three colour stack of the Asiago Schmidt Telescope data. The field of view is approximately $1\times1$\,sq.\,deg.}
    \label{fig:stack}
\end{figure}

\begin{table}
    \centering
    \caption{Log of the observations.}
    \label{tab:obs}
    \begin{tabularx}{\columnwidth}{XXXXX}
         \hline \hline
         Filter & N$_{\rm exp}$ & $t_{\rm exp}$ & Seeing & Airmass \\
         \hline
         $u$ & 57 & 400\,s & 2.25\,arcsec & 1.08 \\
         $g$ & 57 & 400\,s & 2.20\,arcsec & 1.09 \\
         $i$ & 57 & 400\,s & 2.17\,arcsec & 1.07 \\
         \hline
    \end{tabularx}
\end{table}

The data were collected with the Asiago Schmidt telescope between October, 2 and November, 15 2022.
We obtained a set of 57 images in the \textit{Sloan}-like filters $ugi$, with an exposure time of 400\,s.
The images were dithered to mitigate the effect of bad pixels and cosmic rays, and covered a total
area of about 1\,sq.\,deg. The observation log is reported in Table\,\ref{tab:obs}. A three colour stack of
the field of view is shown in Fig\,\ref{fig:stack}, where we used the $u$ filter for the blue colour, $g$ for
the green and $i$ for the red colour.

To measure position and flux of the sources in this dataset we followed the same approach as in \cite{griggio22}. Briefly,
we first derived a grid of $9\times9$ empirical point-spread functions (PSFs) for each image
considering bright, isolated and unsaturated sources,
by using the software originally developed by \cite{2006A&A...454.1029A}.
The grid is necessary to account for spatial variations of the PSF across the CCD. 
We then proceeded by measuring the position and flux of individual sources in each image with the appropriate local PSF, obtained by a bilinear interpolation between the four nearest PSFs in the grid, using the software
described by \cite{2006A&A...454.1029A}.
This routine goes through a series of iterations, finding and measuring progressively fainter sources,
until it reaches a specified level about the sky background noise. The software outputs a catalogue with
positions and instrumental magnitudes for each image.

We transformed the positions and magnitudes of each catalogue to the reference system defined by the first image in each filter (namely, \texttt{SC233176},
\texttt{SC233182} and \texttt{SC233188}).
Finally, we cross identified the sources and produced a catalogue
containing the averaged positions and magnitudes for all the stars
measured in at least five exposures. These catalogues were matched
with the {\it Gaia} one, to have {\it Sloan} magnitudes
for all the {\it Gaia} sources detected with the Schmidt telescope.

The instrumental magnitudes have been calibrated as in \cite{griggio22} 
exploiting the IGAPS catalogue \cite{2020A&A...638A..18M}. We cross identified
our sources with those in the IGAPS catalogue, and derived the coefficients of
the relation $m_{\rm cal}=m_{\rm instr}+a(g_{\rm instr}-i_{\rm instr})+b$ 
with a linear fit.

\begin{figure}
    \centering
    \includegraphics[width=\columnwidth]{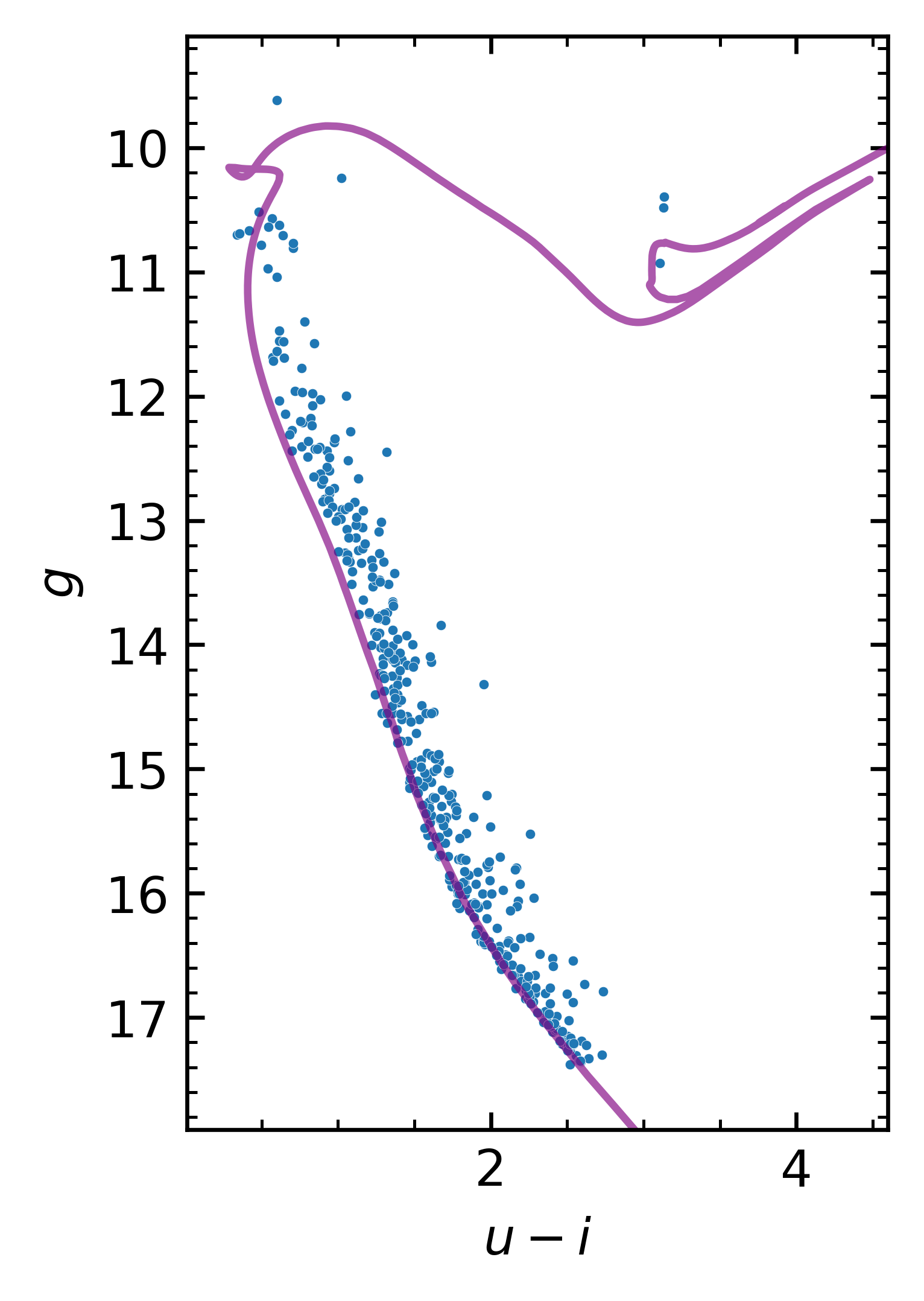}
    \caption{CMD in the {\it Sloan} filters for M38 members. The purple line is the same
    isochrone of Fig.\,\ref{fig:cmd_gaia} (see text for details).}
    \label{fig:cmd_sloan}
\end{figure}

The CMD of member stars in the $ugi$ filters is shown in Fig.\,\ref{fig:cmd_sloan},
together with the same isochrone (purple line) of Fig.\,\ref{fig:cmd_gaia}, employing  
the same distance and reddening, and the 
extinction law from the NASA/IPAC infrared science archive\footnote{\url{https://irsa.ipac.caltech.edu/applications/DUST/}} for the {\it Sloan} filters.

\section{The broadening of the lower MS}
\label{broad}

To investigate in detail the origin of the broadening of the lower MS we followed the same 
technique described in \citet{griggiospread}. We considered stars in the 
{\it Gaia} CMD with $G$ between 15.2 and 16.6
(we have a total of 132 stars in this magnitude range) and combined the photometry in the {\it Gaia} filters with the corresponding $u$ and $i$ magnitudes to build a differential colour-colour diagram, as summarised below.

\begin{figure}
    \centering
    \includegraphics[width=\columnwidth]{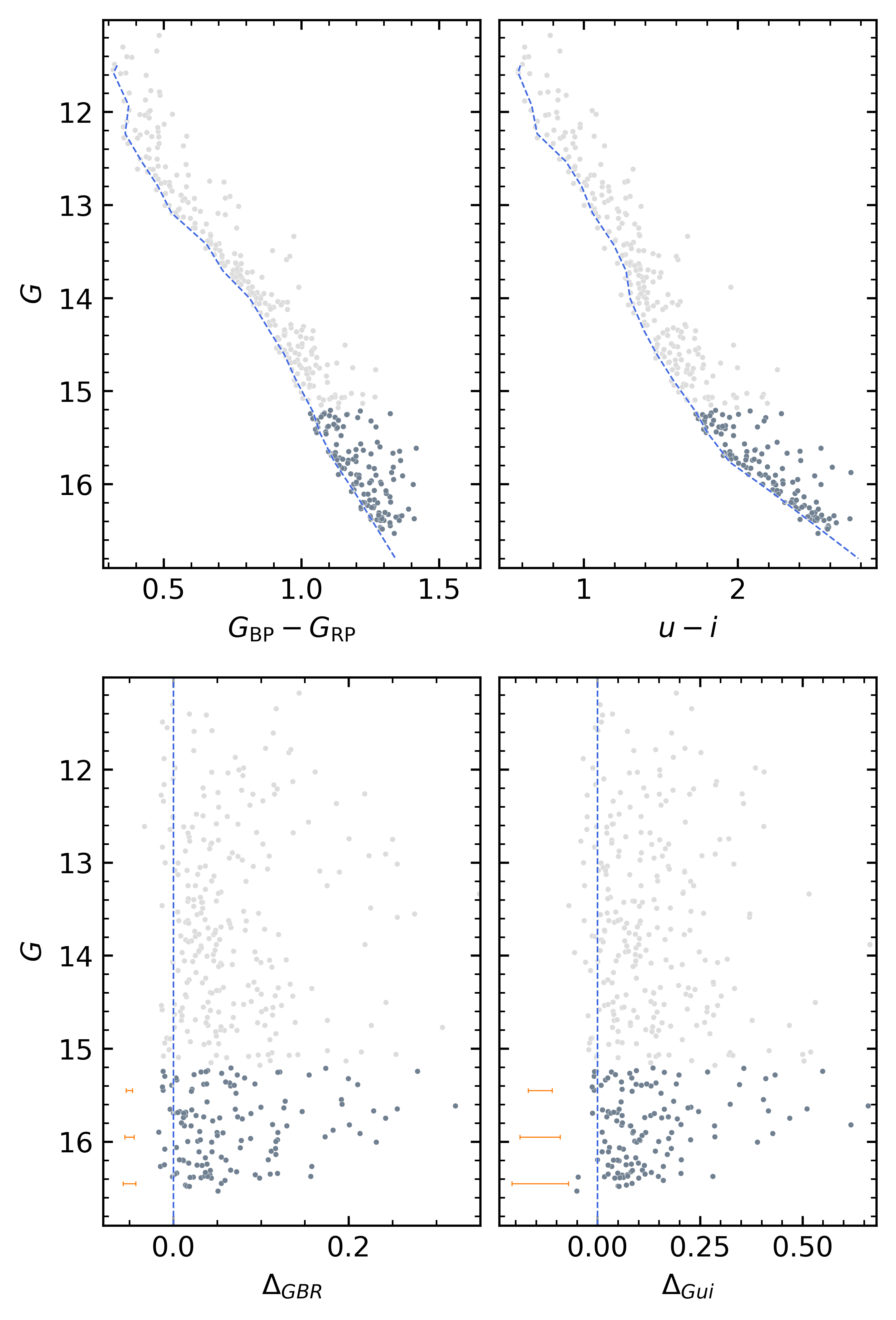}
    \caption{\textit{Top panels}: cluster $G$-$(G_{\rm BP}-G_{\rm RP})$ and $G$-$(u-i)$
    diagrams. The blue dashed line denotes the MS blue edge fiducial. The lower-MS stars considered
    in our analysis are displayed in dark grey. \textit{Bottom panels}: $G$-$\Delta_{GBR}$
    and $G$-$\Delta_{Gui}$ diagrams (see text for details). Along the left side of each panel, 
    we display the median $\pm1\sigma$ colour error for three representative $G$ magnitudes, as estimated
    from the individual catalogues.}
    \label{fig:dd_res}
\end{figure}

We have defined an MS blue fiducial in both the $G$-$(G_{\mathrm{BP}}-G_{\mathrm{RP}})$ and $G$-$(u-i)$ diagrams as described in \citet{griggiospread} and 
for each observed star we have computed, in both $G$-$(G_{\mathrm{BP}}-G_{\mathrm{RP}})$ and $G$-$(u-i)$ diagrams,
the difference between its colour and the colour of the corresponding blue fiducial at the star $G$ magnitude. 
These quantities are denoted as $\Delta_{GBR}$ and $\Delta_{Gui}$ respectively (see Fig.\,\ref{fig:dd_res}).

\begin{figure}
    \centering
    \includegraphics[width=\columnwidth]{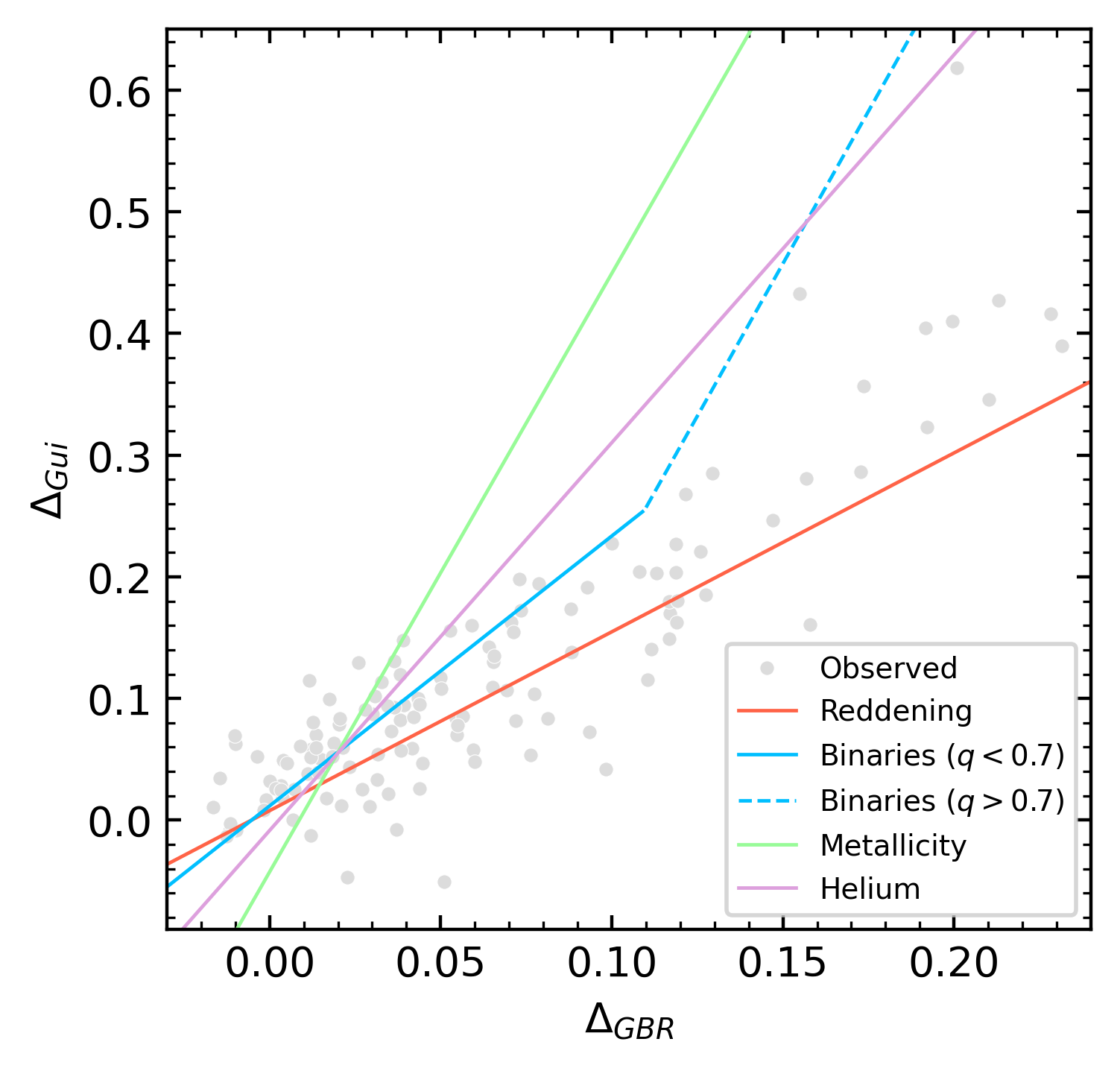}
    \caption{$\Delta_{GBR}$-$\Delta_{Gui}$ diagram for the lower MS stars.
    The lines show the directions along which stars would be displaced by differential reddening,
    unresolved binaries, and initial chemical spread (see text for details).}
    \label{fig:dd}
\end{figure}

We then plotted these colour differences in the 
$\Delta_{GBR}$-$\Delta_{Gui}$ diagram shown 
Fig.\,\ref{fig:dd}. As for the case of M37, 
the lower MS stars are distributed along a well-defined sequence 
which starts around the coordinates (0,0) --corresponding to the  stars lying on the blue fiducials-- and stretches towards 
increasingly positive values (denoting stars increasingly redder than the fiducials) with the quantity $\Delta_{Gui}$ increasing faster than $\Delta_{GBR}$.
These colour spreads cannot arise from (underestimated) random photometric 
errors only, because in this case they would be distributed without a correlation 
between $\Delta_{Gui}$ and $\Delta_{GBR}$. 

In the same figure, together with the data, we show  
the reddening vector, calculated using the extinction laws for the {\it Gaia} 
and {\it Sloan } filters referenced above.
We also plot the vector corresponding to the predicted position 
of binaries with varying mass ratio $q$ (blue) and the range of colours spanned by 
isochrones with increasing [Fe/H] and increasing $Y$ (green and magenta).
These vector have been calculated as described 
by \citet{griggiospread} for M37, 
using as reference the isochrone in Fig.\,\ref{fig:cmd_gaia}, and  
the corresponding values of $d$ and $E(B-V)$ (see Sect.\,\ref{Gaia}).
In this figure, we display the effect of binaries as a two-slope  
sequence, because it is a better representation of the trend predicted by synthetic stellar populations, compared to a single slope as shown in \citet{griggiospread}.

The figure shows that, in the case of M38, the distribution of the stars' colours in this diagram follows a trend consistent with a combination of 
differential reddening across the face of the cluster and the presence of unresolved binaries with varying $q$. There is no need to invoke 
the presence of a range of [Fe/H] or $Y$ among the clusters' stars.
This is at odds with the case of M37, where binaries and differential reddening produced too shallow slopes in this diagram, compared to the observations (see the Appendix for a comparison of the CMDs and $\Delta_{GBR}$-$\Delta_{Gui}$ diagrams of M37 and M38).

\begin{figure}
    \centering
    \includegraphics[width=\columnwidth]{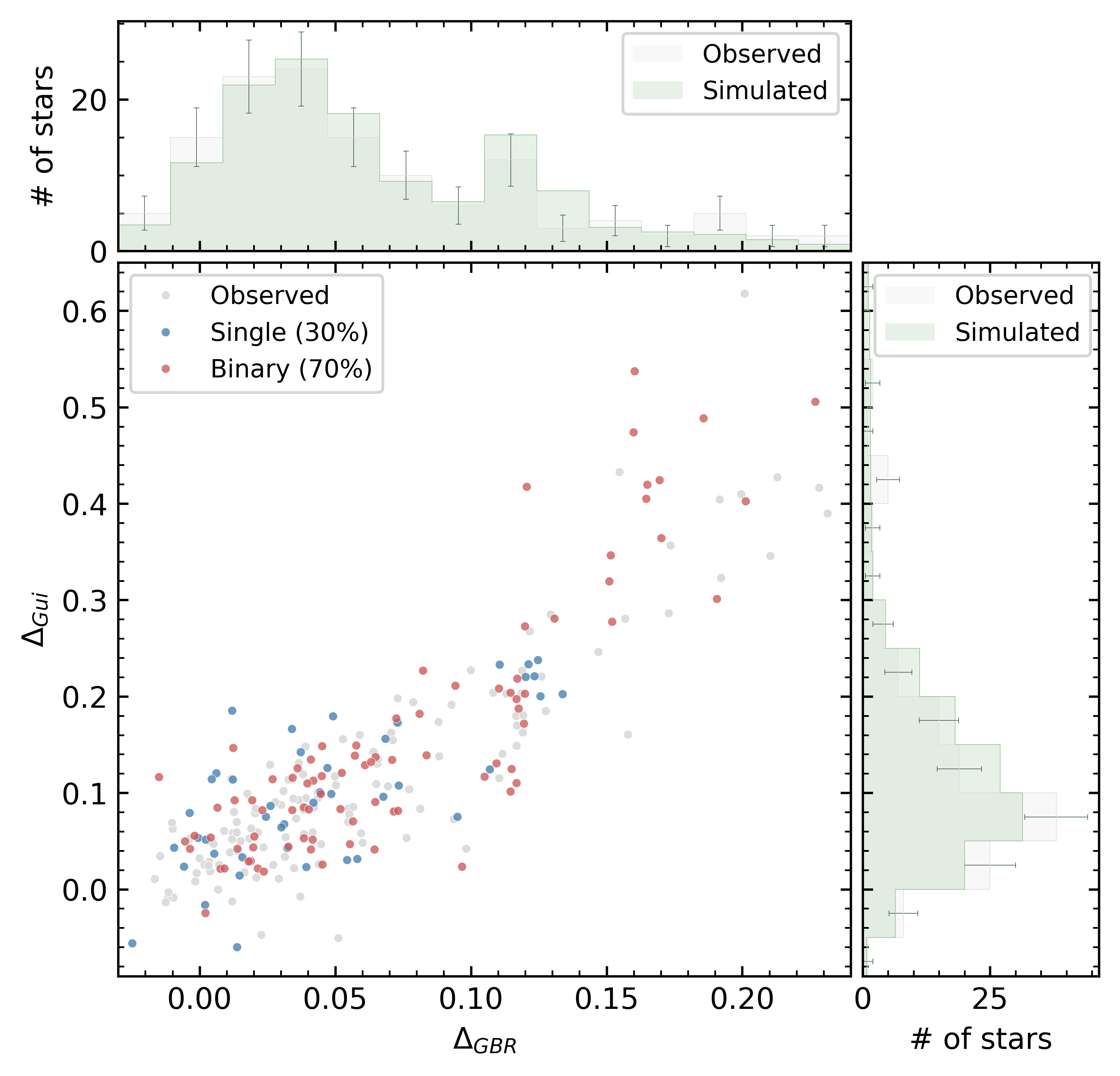}
    \caption{Comparison of the distribution of the synthetic population (including binaries and a spread of $E(B-V)$ to account for differential reddening -- see text for details)
    and the observed cluster stars (grey filled circles) in the 
    $\Delta_{GBR}$-$\Delta_{Gui}$ diagram. 
    Synthetic single and unresolved binary stars are shown in blue and red respectively.
    Histograms of the observed (with Poisson error bars) 
    and simulated number counts along the two axes are also shown (see text for details).}
    \label{fig:dd_bin}
\end{figure}

Figure\,\ref{fig:dd_bin} shows a synthetic sample of stars --computed as in \citet{griggiospread}, using isochrone, distance, and reference reddening previously discussed-- compared to observations in the $\Delta_{GBR}$-$\Delta_{Gui}$ diagram. The purpose of this comparison is just to see how binaries and differential reddening only can account qualitatively for the observed distribution of lower-MS stars in 
this diagram.

The full synthetic sample of 50\,000 objects includes observational errors in both the {\it Gaia}
and {\it Sloan} magnitudes, and contains a 70\,\% fraction of unresolved binaries with mass-ratios $q$ distributed as 
$f(q) \propto q^{-0.6}$ following \cite{malofeeva}. It is worth pointing out that in the case of assuming a flat probability distribution for $q$, the same results described below are obtained with a 10-15\,\% binary fraction. 

We display here one random subset of the full sample, containing the same number of objects as the observations. 
The figure also shows 
along the horizontal and vertical axis a comparison of the number distributions of synthetic and observed stars as a function of the two quantities $\Delta_{GBR}$ and $\Delta_{Gui}$, 
respectively. When calculating these histograms we have considered the full sample of synthetic stars and rescaled the derived histograms to have the same total number of objects as observed.

The contribution of differential reddening has been accounted for by using a double Gaussian distribution;
only in this way, we are able to reproduce the clump of stars clearly visible in the $\Delta_{GBR}$ histogram at $\Delta_{GBR}\,\sim$\,0.12.
The parameters of the distributions have been adjusted to roughly reproduce the observed trends of the number distributions in both $\Delta_{GBR}$ and $\Delta_{Gui}$, because we could not determine 
a reliable differential reddening map for M38 using the technique described by, e.g.,
\cite{diffred}, given the relatively low number of objects. The 
first Gaussian accounts for a random sample of $\sim$\,80\,\% of the synthetic 
stars (both single and binary objects) and 
is centred on $E(B-V)=E(B-V)_{\rm ref}+0.04$, with $\sigma=0.033$\,mag, where $E(B-V)_{\rm ref}=0.26$. The second Gaussian distribution is centred on $E(B-V)=E(B-V)_{\rm ref}+0.15$, with $\sigma=0.01$\,mag, and accounts for the remaining objects in the synthetic sample.

It is remarkable how this synthetic sample, which includes just unresolved binaries and the effect of differential reddening, follows nicely the observed trend in this diagram. Also, the observed number distribution across the diagram can be followed quite well by using two simple Gaussian 
$E(B-V)$ distributions and the power-law $q$ distribution determined 
by \citet{malofeeva}. 
This shows that there is no need to invoke a chemical abundance spread to explain the width of the lower MS in this cluster.

We have then repeated the analysis previously described considering 
this time stars in the brighter $G$ magnitude range between 12.5 and 14, corresponding to single star masses between $\sim$\,1.5 and $\sim$\,2.2\,$M_{\odot}$. 
Using the same binary fraction, $q$ and $E(B-V)$ distributions of the previous comparison, we have found the same agreement of the number distributions of synthetic and observed stars across the $\Delta_{GBR}$-$\Delta_{Gui}$ diagram as in Fig.,\ref{fig:dd_bin}. 

The same comparison could not be performed for objects in the TO region of the CMD, because of the small sample of cluster stars in this magnitude range (see Sect.\,\ref{Gaia}).

\section{Summary and conclusions}
\label{conclusions}

We have employed the accurate {\it Gaia} DR3 photometry and astrometry 
of the poorly studied open cluster M38 to select {\it bona fide} members and determine the cluster distance and mean proper motion.
The {\it Gaia} CMD does not show an obvious extended TO despite the  
cluster being $\sim$\,300\,Myr old, but the number of stars in the TO region is too small to investigate quantitatively this matter.
The unevolved MS is broader than expected from photometric errors only 
and to determine the origin of this broadening we have applied 
the same technique developed to study the open cluster M37 \citep{griggiospread}, making use of auxiliary photometry in the {\it Sloan} system to build a differential colour-colour diagram of the lower MS from combinations of {\it Gaia} and {\it Sloan} magnitudes.

We employed synthetic stellar populations to reproduce the observed trend of M38 stars in this diagram, and found that 
the observed MS colour spread can be explained simply by the combined effect 
of differential reddening and unresolved binaries.
There is no need to include a spread of initial chemical composition (either metals or helium) as instead necessary to explain the same differential colour-colour diagram for the lower MS of M37.

Despite having similar total masses 
\citep[estimated total masses on the order of 1\,000-1\,500\,$M_{\odot}$, mutually consistent within the associated errors, see][]{piskunov} 
and metallicities different on average by no more than at most a 
factor 2-3, the open clusters M38 and M37 seem to host stellar populations with a clear difference: single vs multiple chemical 
compositions.
The origin of this difference is unknown and we do not know as well whether the chemical abundance spread found photometrically in M37 is 
a feature common to many more open clusters, and if there is any connection with the extended TO phenomenon.

Further photometric investigations like ours, as well as accurate differential spectroscopic analyses on a large sample 
of open clusters are necessary to shed light on this phenomenon, and its implications for cluster formation and the use of open clusters and chemical tagging to study the formation and evolution of the Galactic disk.

\section*{Acknowledgements}
Based on observations collected at the Schmidt telescope (Asiago, Italy) of INAF.

This work has made use of data from the European Space Agency (ESA) mission
{\it Gaia} (\url{https://www.cosmos.esa.int/gaia}), processed by the {\it Gaia}
Data Processing and Analysis Consortium (DPAC,
\url{https://www.cosmos.esa.int/web/gaia/dpac/consortium}). Funding for the DPAC
has been provided by national institutions, in particular the institutions
participating in the {\it Gaia} Multilateral Agreement.

This research or product makes use of public auxiliary data provided by ESA/Gaia/DPAC/CU5 
and prepared by Carine Babusiaux.

MG and LRB acknowledge support by MIUR under PRIN program \#2017Z2HSMF and PRIN INAF 2019 (PI Bedin). 

MS acknowledges support from The Science and
Technology Facilities Council Consolidated Grant ST/V00087X/1.

SC acknowledges financial support from Premiale INAF MITiC, from INFN (Iniziativa specifica TAsP), and from
PLATO ASI-INAF agreement n.2015-019-R.1-2018.

\section*{Data Availability}

The isochrones employed in this study can be retrieved at \url{http://basti-iac.oa-abruzzo.inaf.it}, 
but for the helium enhanced isochrones, that are available upon request. 

The calibrated photometry and astrometry employed in this article is released
as supplementary on-line material, and available at 
\url{https://web.oapd.inaf.it/bedin/files/PAPERs\_eMATERIALs/M38\_ugiSchmidt/}, 
along with an atlas.
The same catalogue also conveniently lists the \textit{Gaia}\,DR3 photometry, astrometry and 
source ID, when available.



\bibliographystyle{mnras}
\bibliography{bibliography} 

\begin{thebibliography}{}
\makeatletter
\relax
\def\mn@urlcharsother{\let\do\@makeother \do\$\do\&\do\#\do\^\do\_\do\%\do\~}
\def\mn@doi{\begingroup\mn@urlcharsother \@ifnextchar [ {\mn@doi@}
  {\mn@doi@[]}}
\def\mn@doi@[#1]#2{\def\@tempa{#1}\ifx\@tempa\@empty \href
  {http://dx.doi.org/#2} {doi:#2}\else \href {http://dx.doi.org/#2} {#1}\fi
  \endgroup}
\def\mn@eprint#1#2{\mn@eprint@#1:#2::\@nil}
\def\mn@eprint@arXiv#1{\href {http://arxiv.org/abs/#1} {{\tt arXiv:#1}}}
\def\mn@eprint@dblp#1{\href {http://dblp.uni-trier.de/rec/bibtex/#1.xml}
  {dblp:#1}}
\def\mn@eprint@#1:#2:#3:#4\@nil{\def\@tempa {#1}\def\@tempb {#2}\def\@tempc
  {#3}\ifx \@tempc \@empty \let \@tempc \@tempb \let \@tempb \@tempa \fi \ifx
  \@tempb \@empty \def\@tempb {arXiv}\fi \@ifundefined
  {mn@eprint@\@tempb}{\@tempb:\@tempc}{\expandafter \expandafter \csname
  mn@eprint@\@tempb\endcsname \expandafter{\@tempc}}}

\bibitem[\protect\citeauthoryear{{Anderson}, {Bedin}, {Piotto}, {Yadav}  \&
  {Bellini}}{{Anderson} et~al.}{2006}]{2006A&A...454.1029A}
{Anderson} J.,  {Bedin} L.~R.,  {Piotto} G.,  {Yadav} R.~S.,   {Bellini} A.,
  2006, \mn@doi [\aap] {10.1051/0004-6361:20065004}, \href
  {https://ui.adsabs.harvard.edu/abs/2006A&A...454.1029A} {454, 1029}

\bibitem[\protect\citeauthoryear{{Bailer-Jones}, {Rybizki}, {Fouesneau},
  {Demleitner}  \& {Andrae}}{{Bailer-Jones} et~al.}{2021}]{2021AJ....161..147B}
{Bailer-Jones} C.~A.~L.,  {Rybizki} J.,  {Fouesneau} M.,  {Demleitner} M.,
  {Andrae} R.,  2021, \mn@doi [\aj] {10.3847/1538-3881/abd806}, \href
  {https://ui.adsabs.harvard.edu/abs/2021AJ....161..147B} {161, 147}

\bibitem[\protect\citeauthoryear{{Bastian}, {Kamann}, {Cabrera-Ziri}, {Georgy},
  {Ekstr{\"o}m}, {Charbonnel}, {de Juan Ovelar}  \& {Usher}}{{Bastian}
  et~al.}{2018}]{bastian}
{Bastian} N.,  {Kamann} S.,  {Cabrera-Ziri} I.,  {Georgy} C.,  {Ekstr{\"o}m}
  S.,  {Charbonnel} C.,  {de Juan Ovelar} M.,   {Usher} C.,  2018, \mn@doi
  [\mnras] {10.1093/mnras/sty2100}, \href
  {https://ui.adsabs.harvard.edu/abs/2018MNRAS.480.3739B} {480, 3739}

\bibitem[\protect\citeauthoryear{{Carrera, R.} et~al.,}{{Carrera, R.}
  et~al.}{2019}]{carrera}
{Carrera, R.} et~al., 2019, \mn@doi [A\&A] {10.1051/0004-6361/201834546}, 623,
  A80

\bibitem[\protect\citeauthoryear{{Clarke}, {Bonnell}  \&
  {Hillenbrand}}{{Clarke} et~al.}{2000}]{clarke}
{Clarke} C.~J.,  {Bonnell} I.~A.,   {Hillenbrand} L.~A.,  2000, in {Mannings}
  V.,  {Boss} A.~P.,   {Russell} S.~S.,  eds, Protostars and Planets IV. p.~151
  (\mn@eprint {arXiv} {astro-ph/9903323}),
  \mn@doi{10.48550/arXiv.astro-ph/9903323}

\bibitem[\protect\citeauthoryear{{Cordoni}, {Milone}, {Marino}, {Di
  Criscienzo}, {D'Antona}, {Dotter}, {Lagioia}  \& {Tailo}}{{Cordoni}
  et~al.}{2018}]{cordoni18}
{Cordoni} G.,  {Milone} A.~P.,  {Marino} A.~F.,  {Di Criscienzo} M.,
  {D'Antona} F.,  {Dotter} A.,  {Lagioia} E.~P.,   {Tailo} M.,  2018, \mn@doi
  [\apj] {10.3847/1538-4357/aaedc1}, \href
  {https://ui.adsabs.harvard.edu/abs/2018ApJ...869..139C} {869, 139}

\bibitem[\protect\citeauthoryear{{Correnti}, {Goudfrooij}, {Kalirai},
  {Girardi}, {Puzia}  \& {Kerber}}{{Correnti} et~al.}{2014}]{corr}
{Correnti} M.,  {Goudfrooij} P.,  {Kalirai} J.~S.,  {Girardi} L.,  {Puzia}
  T.~H.,   {Kerber} L.,  2014, \mn@doi [\apj] {10.1088/0004-637X/793/2/121},
  \href {https://ui.adsabs.harvard.edu/abs/2014ApJ...793..121C} {793, 121}

\bibitem[\protect\citeauthoryear{{D'Antona} et~al.,}{{D'Antona}
  et~al.}{2023}]{dantona}
{D'Antona} F.,  et~al., 2023, \mn@doi [arXiv e-prints]
  {10.48550/arXiv.2303.16049}, \href
  {https://ui.adsabs.harvard.edu/abs/2023arXiv230316049D} {p. arXiv:2303.16049}

\bibitem[\protect\citeauthoryear{{Dias}, {Alessi}, {Moitinho}  \&
  {L{\'e}pine}}{{Dias} et~al.}{2002}]{dias}
{Dias} W.~S.,  {Alessi} B.~S.,  {Moitinho} A.,   {L{\'e}pine} J.~R.~D.,  2002,
  \mn@doi [\aap] {10.1051/0004-6361:20020668}, \href
  {https://ui.adsabs.harvard.edu/abs/2002A&A...389..871D} {389, 871}

\bibitem[\protect\citeauthoryear{{Donor} et~al.,}{{Donor} et~al.}{2020}]{donor}
{Donor} J.,  et~al., 2020, \mn@doi [\aj] {10.3847/1538-3881/ab77bc}, \href
  {https://ui.adsabs.harvard.edu/abs/2020AJ....159..199D} {159, 199}

\bibitem[\protect\citeauthoryear{{Freeman} \& {Bland-Hawthorn}}{{Freeman} \&
  {Bland-Hawthorn}}{2002}]{tagging}
{Freeman} K.,  {Bland-Hawthorn} J.,  2002, \mn@doi [\araa]
  {10.1146/annurev.astro.40.060401.093840}, \href
  {https://ui.adsabs.harvard.edu/abs/2002ARA&A..40..487F} {40, 487}

\bibitem[\protect\citeauthoryear{{Frinchaboy} et~al.,}{{Frinchaboy}
  et~al.}{2013}]{frincha}
{Frinchaboy} P.~M.,  et~al., 2013, \mn@doi [\apjl]
  {10.1088/2041-8205/777/1/L1}, \href
  {https://ui.adsabs.harvard.edu/abs/2013ApJ...777L...1F} {777, L1}

\bibitem[\protect\citeauthoryear{{Gaia Collaboration} et~al.,}{{Gaia
  Collaboration} et~al.}{2022}]{2022arXiv220800211G}
{Gaia Collaboration} et~al., 2022, \mn@doi [arXiv e-prints]
  {10.48550/arXiv.2208.00211}, \href
  {https://ui.adsabs.harvard.edu/abs/2022arXiv220800211G} {p. arXiv:2208.00211}

\bibitem[\protect\citeauthoryear{{Goudfrooij}, {Girardi}  \&
  {Correnti}}{{Goudfrooij} et~al.}{2017}]{goudrot}
{Goudfrooij} P.,  {Girardi} L.,   {Correnti} M.,  2017, \mn@doi [\apj]
  {10.3847/1538-4357/aa7fb7}, \href
  {https://ui.adsabs.harvard.edu/abs/2017ApJ...846...22G} {846, 22}

\bibitem[\protect\citeauthoryear{{Griggio} \& {Bedin}}{{Griggio} \&
  {Bedin}}{2022}]{gb2022}
{Griggio} M.,  {Bedin} L.~R.,  2022, \mn@doi [\mnras] {10.1093/mnras/stac391},
  \href {https://ui.adsabs.harvard.edu/abs/2022MNRAS.511.4702G} {511, 4702}

\bibitem[\protect\citeauthoryear{{Griggio} et~al.,}{{Griggio}
  et~al.}{2022a}]{griggio22}
{Griggio} M.,  et~al., 2022a, \mn@doi [\mnras] {10.1093/mnras/stac1920}, \href
  {https://ui.adsabs.harvard.edu/abs/2022MNRAS.515.1841G} {515, 1841}

\bibitem[\protect\citeauthoryear{{Griggio}, {Salaris}, {Cassisi},
  {Pietrinferni}  \& {Bedin}}{{Griggio} et~al.}{2022b}]{griggiospread}
{Griggio} M.,  {Salaris} M.,  {Cassisi} S.,  {Pietrinferni} A.,   {Bedin}
  L.~R.,  2022b, \mn@doi [\mnras] {10.1093/mnras/stac2512}, \href
  {https://ui.adsabs.harvard.edu/abs/2022MNRAS.516.3631G} {516, 3631}

\bibitem[\protect\citeauthoryear{{Hidalgo} et~al.,}{{Hidalgo}
  et~al.}{2018}]{hidalgo18}
{Hidalgo} S.~L.,  et~al., 2018, \mn@doi [\apj] {10.3847/1538-4357/aab158},
  \href {https://ui.adsabs.harvard.edu/abs/2018ApJ...856..125H} {856, 125}

\bibitem[\protect\citeauthoryear{{Hogg} et~al.,}{{Hogg} et~al.}{2016}]{hogg}
{Hogg} D.~W.,  et~al., 2016, \mn@doi [\apj] {10.3847/1538-4357/833/2/262},
  \href {https://ui.adsabs.harvard.edu/abs/2016ApJ...833..262H} {833, 262}

\bibitem[\protect\citeauthoryear{{Kamann} et~al.,}{{Kamann}
  et~al.}{2018}]{kamann18}
{Kamann} S.,  et~al., 2018, \mn@doi [\mnras] {10.1093/mnras/sty1958}, \href
  {https://ui.adsabs.harvard.edu/abs/2018MNRAS.480.1689K} {480, 1689}

\bibitem[\protect\citeauthoryear{{Kamann} et~al.,}{{Kamann}
  et~al.}{2020}]{kamann20}
{Kamann} S.,  et~al., 2020, \mn@doi [\mnras] {10.1093/mnras/stz3583}, \href
  {https://ui.adsabs.harvard.edu/abs/2020MNRAS.492.2177K} {492, 2177}

\bibitem[\protect\citeauthoryear{{Kamann} et~al.,}{{Kamann}
  et~al.}{2023}]{kamann23}
{Kamann} S.,  et~al., 2023, \mn@doi [\mnras] {10.1093/mnras/stac3170}, \href
  {https://ui.adsabs.harvard.edu/abs/2023MNRAS.518.1505K} {518, 1505}

\bibitem[\protect\citeauthoryear{{Li} et~al.,}{{Li} et~al.}{2021}]{li}
{Li} C.-Y.,  et~al., 2021, \mn@doi [Research in Astronomy and Astrophysics]
  {10.1088/1674-4527/21/3/068}, \href
  {https://ui.adsabs.harvard.edu/abs/2021RAA....21...68L} {21, 068}

\bibitem[\protect\citeauthoryear{{Lindegren} et~al.,}{{Lindegren}
  et~al.}{2021}]{2021A&A...649A...4L}
{Lindegren} L.,  et~al., 2021, \mn@doi [\aap] {10.1051/0004-6361/202039653},
  \href {https://ui.adsabs.harvard.edu/abs/2021A&A...649A...4L} {649, A4}

\bibitem[\protect\citeauthoryear{{Mackey} \& {Broby Nielsen}}{{Mackey} \&
  {Broby Nielsen}}{2007}]{mb}
{Mackey} A.~D.,  {Broby Nielsen} P.,  2007, \mn@doi [\mnras]
  {10.1111/j.1365-2966.2007.11915.x}, \href
  {https://ui.adsabs.harvard.edu/abs/2007MNRAS.379..151M} {379, 151}

\bibitem[\protect\citeauthoryear{{Mackey}, {Broby Nielsen}, {Ferguson}  \&
  {Richardson}}{{Mackey} et~al.}{2008}]{mackey}
{Mackey} A.~D.,  {Broby Nielsen} P.,  {Ferguson} A.~M.~N.,   {Richardson}
  J.~C.,  2008, \mn@doi [\apjl] {10.1086/590343}, \href
  {https://ui.adsabs.harvard.edu/abs/2008ApJ...681L..17M} {681, L17}

\bibitem[\protect\citeauthoryear{{Majaess}, {Turner}  \& {Lane}}{{Majaess}
  et~al.}{2007}]{majaess}
{Majaess} D.~J.,  {Turner} D.~G.,   {Lane} D.~J.,  2007, \mn@doi [\pasp]
  {10.1086/524414}, \href
  {https://ui.adsabs.harvard.edu/abs/2007PASP..119.1349M} {119, 1349}

\bibitem[\protect\citeauthoryear{{Malofeeva}, {Seleznev}  \&
  {Carraro}}{{Malofeeva} et~al.}{2022}]{malofeeva}
{Malofeeva} A.~A.,  {Seleznev} A.~F.,   {Carraro} G.,  2022, \mn@doi [\aj]
  {10.3847/1538-3881/ac47a3}, \href
  {https://ui.adsabs.harvard.edu/abs/2022AJ....163..113M} {163, 113}

\bibitem[\protect\citeauthoryear{{Marino}, {Milone}, {Casagrande}, {Przybilla},
  {Balaguer-N{\'u}{\~n}ez}, {Di Criscienzo}, {Serenelli}  \&
  {Vilardell}}{{Marino} et~al.}{2018}]{marino18}
{Marino} A.~F.,  {Milone} A.~P.,  {Casagrande} L.,  {Przybilla} N.,
  {Balaguer-N{\'u}{\~n}ez} L.,  {Di Criscienzo} M.,  {Serenelli} A.,
  {Vilardell} F.,  2018, \mn@doi [\apjl] {10.3847/2041-8213/aad868}, \href
  {https://ui.adsabs.harvard.edu/abs/2018ApJ...863L..33M} {863, L33}

\bibitem[\protect\citeauthoryear{{Mongui{\'o}} et~al.,}{{Mongui{\'o}}
  et~al.}{2020}]{2020A&A...638A..18M}
{Mongui{\'o}} M.,  et~al., 2020, \mn@doi [\aap] {10.1051/0004-6361/201937333},
  \href {https://ui.adsabs.harvard.edu/abs/2020A&A...638A..18M} {638, A18}

\bibitem[\protect\citeauthoryear{{Piatti} \& {Bastian}}{{Piatti} \&
  {Bastian}}{2016}]{pb16}
{Piatti} A.~E.,  {Bastian} N.,  2016, \mn@doi [\aap]
  {10.1051/0004-6361/201628339}, \href
  {https://ui.adsabs.harvard.edu/abs/2016A&A...590A..50P} {590, A50}

\bibitem[\protect\citeauthoryear{{Piskunov, A. E.}, {Schilbach, E.},
  {Kharchenko, N. V.}, {R\"oser, S.}  \& {Scholz, R.-D.}}{{Piskunov, A. E.}
  et~al.}{2008}]{piskunov}
{Piskunov, A. E.} {Schilbach, E.} {Kharchenko, N. V.} {R\"oser, S.}  {Scholz,
  R.-D.} 2008, \mn@doi [A\&A] {10.1051/0004-6361:20078525}, 477, 165

\bibitem[\protect\citeauthoryear{{Sarajedini} et~al.,}{{Sarajedini}
  et~al.}{2007}]{diffred}
{Sarajedini} A.,  et~al., 2007, \mn@doi [\aj] {10.1086/511979}, \href
  {https://ui.adsabs.harvard.edu/abs/2007AJ....133.1658S} {133, 1658}

\bibitem[\protect\citeauthoryear{{Subramaniam} \& {Sagar}}{{Subramaniam} \&
  {Sagar}}{1999}]{subramaniam}
{Subramaniam} A.,  {Sagar} R.,  1999, \mn@doi [\aj] {10.1086/300716}, \href
  {https://ui.adsabs.harvard.edu/abs/1999AJ....117..937S} {117, 937}

\bibitem[\protect\citeauthoryear{{Tarricq} et~al.,}{{Tarricq}
  et~al.}{2021}]{tarricq}
{Tarricq} Y.,  et~al., 2021, \mn@doi [\aap] {10.1051/0004-6361/202039388},
  \href {https://ui.adsabs.harvard.edu/abs/2021A&A...647A..19T} {647, A19}

\bibitem[\protect\citeauthoryear{{Zhong}, {Chen}, {Wu}, {Li}, {Bai}  \&
  {Hou}}{{Zhong} et~al.}{2020}]{zhong}
{Zhong} J.,  {Chen} L.,  {Wu} D.,  {Li} L.,  {Bai} L.,   {Hou} J.,  2020,
  \mn@doi [\aap] {10.1051/0004-6361/201937131}, \href
  {https://ui.adsabs.harvard.edu/abs/2020A&A...640A.127Z} {640, A127}

\makeatother
\end{thebibliography}



\appendix

\section{Comparison of M37 and M38 diagrams}
\begin{figure}
    \centering
    \includegraphics[width=\columnwidth]{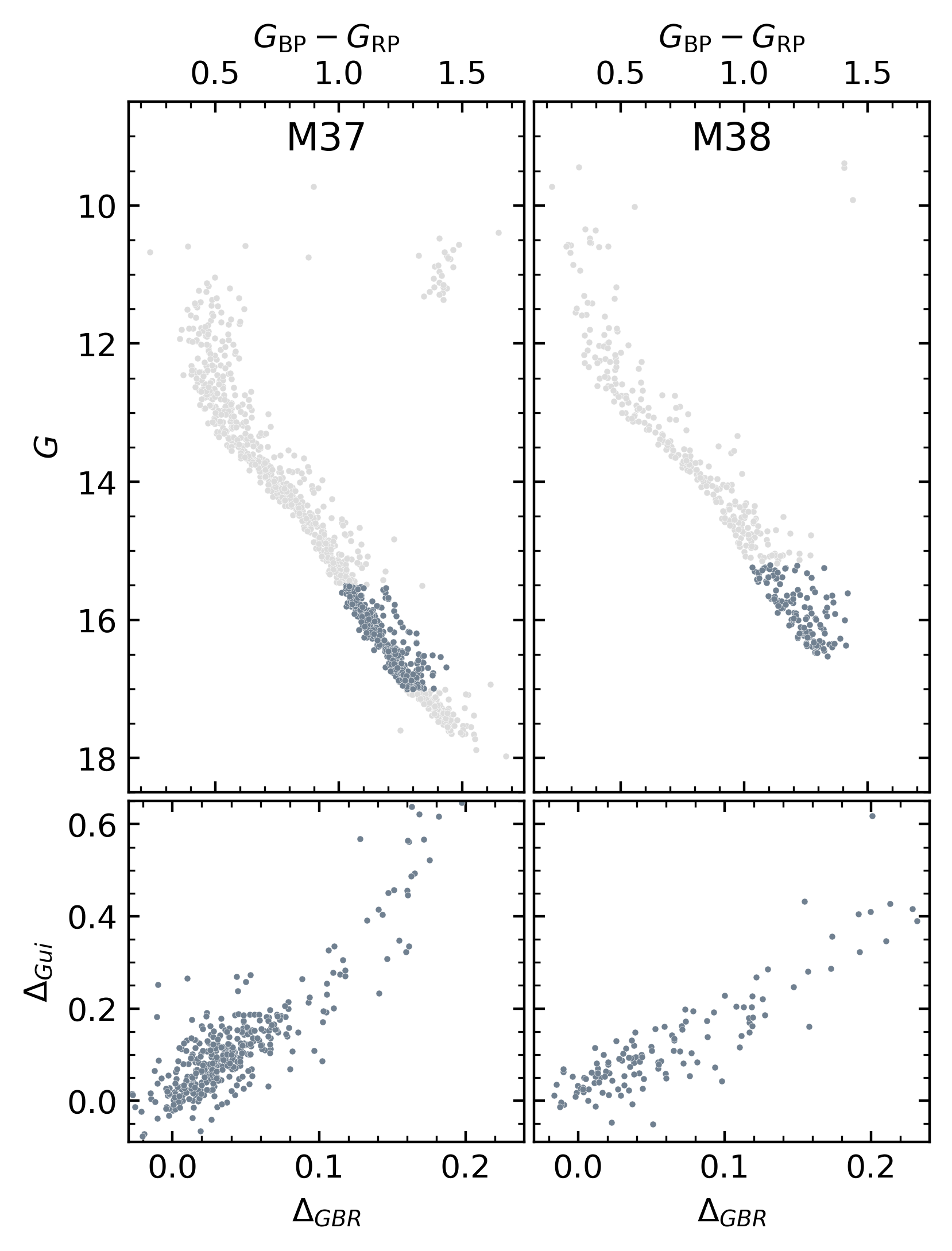}
    \caption{Comparison of M37 and M38 CMDs and $\Delta_{GBR}$-$\Delta_{Gui}$ 
    diagrams.}
    \label{fig:compare}
\end{figure}

We present here a comparison of the {\it Gaia} CMD and the 
$\Delta_{GBR}$-$\Delta_{Gui}$ diagrams of M37 and M38. The top panels
of Fig.\,\ref{fig:compare} show the CMDs of our M37 (left) and M38 (right) {\sl bona-fide} members, highlighting in a darker shade of grey 
the sources employed in the analysis of the width of the MS. The bottom panels display the $\Delta_{GBR}$-$\Delta_{Gui}$
diagrams of both cluster's lower MS. The points are clearly distributed along different
slopes in the two diagrams; in particular, we also notice the behaviour of high-$q$ binaries
($\Delta_{GBR}\gtrsim 0.1$) in the case of M37, that are distributed along a steeper
line, while M38 stars in the same region follow a shallower direction.

\bsp	
\label{lastpage}
\end{document}